\documentclass[11pt]{article}
\usepackage{amssymb}
\usepackage{amsmath}
\usepackage{graphicx}
\usepackage{epsfig}
\usepackage{amsfonts}
\usepackage{color}
\usepackage{float}
\usepackage{latexsym}
\usepackage{mathrsfs}
\usepackage{feynmf}
\usepackage{balance}

\def\be{\begin{equation}} \def\ee{\end{equation}}
\def\ba{\begin{eqnarray}} \def\ea{\end{eqnarray}} \def\part{\partial}

   \def\b1{{\bf 1}}

\begin{document}

\begin{titlepage}
\title{
\begin{flushright}\begin{small}
\end{small}\end{flushright}\vspace{2cm}
Collective-Coordinate Analysis of Inhomogeneous Nonlinear Klein-Gordon Field Theory}

\author{Danial Saadatmand\thanks{Email: Da$\_$se.saadatmand.643@stu-mail.um.ac.ir} $\,$ and
Kurosh Javidan\thanks{Email: Javidan@um.ac.ir} \\ \\
\small{Department of Physics, Ferdowsi University of Mashhad,}\\
\small{91775-1436  Mashhad, Iran} }
\maketitle

\abstract{Two different sets of collective-coordinate equations for solitary solutions of nonlinear Klein-Gordon (NKG) model are introduced. The collective-coordinate equations are derived using different approaches for adding the inhomogeneities as external potentials to the soliton equation of motion.  Interaction of the NKG field with a local inhomogeneity like a delta function potential wall as well as a delta function potential well is investigated using the presented collective-coordinate equations and the results of the two different models are compared. Most of the characters of the interaction are derived analytically. Analytical results are also compared to the results of numerical simulations.}
\end{titlepage}
\setcounter{page}{2}
\section{Introduction}
\quad\, Solitons are localized waves that have a nonzero energy density in a finite region of space which exponentially goes to zero as one moves away from this region. They appear in nonlinear classical field theories as stable and particle-like objects with finite mass and explicit structures. Therefore, finding suitable methods for studying the soliton as a point particle helps us to find a better perspective of the soliton behaviour. On the other hand, comparing the results of such kinds of models with the results of direct numerical simulations determines the differences between solitons as point-like particles and real solitons. This topic is an interesting subject in nonlinear field theories \cite{i1}. Solitons appear in a nonlinear medium with a fine tuning between nonlinear and dispersive effects. This means that they may disappear in the absence of this precise balance in the medium. It is clear that a real medium contains disorders and impurities. Therefore, stability and propagation of solitons in such media are of great interest because of their applications and theoretical interests. In order to understand the behaviour of nonlinear excitations in a disordered system, it is important to investigate the interaction of  solitons with impurities.

Recently, some non-classical behaviours have been reported for solitons during the scattering from external potentials \cite{i2}. These potentials are generally due to medium defects or impurities. The scattering of solitons of integrable systems from the potentials have been studied before\cite{i3}; but such an investigation for non integrable systems has not been reported yet. Therefore, it is interesting to examine the methods of adding the potential to the NKG model as a non-integrable model and compare the results to those of integrable systems. These are strong motivations for investigating the scattering of the NKG solitons from defects.

External potentials can be added to the equation of motion using different methods. One way to do it is to add an external potential to the equation of motion as perturbative terms \cite{i2,i3}. These effects can also be taken into account by making some parameters of the equation of motion to become a function of space or time \cite{i4,i5}. Another way to indicate it is by adding an external potential to the field through the metric of background space-time \cite{i6, i6a, i7}. This method can be used for models in which the Lagrangians are Lorentz invariant, such as Sine-Gordon model, $ \phi^{4} $ theory, $CP^N$ model, NKG models, etc. In this paper we will focus on the behaviour of solitons of the NKG and try to investigate the interaction of the NKG solitons with defects by means of two different analytical models.

Different types of the NKG models, which appear in some branches of science, are  important non-integrable models. These equations can be used to describe the particle dynamics in quantum field theory. Some of the other examples of the NKG applications include, discrete gap breathers in a diatomic chain \cite{i8}, dichotomous collective proton dynamics in ice \cite{i9}, propagation and stability of relaxation modes in the Landau-Ginzburg model with dissipation\cite{i10} and pion form factor \cite{i11}. Recently, Wazwaz has proposed several localized solutions for the NKG equations using "Tanh" method \cite{i12}. Solitons present different trajectories during the interaction with potentials. They can either pass through or become trapped inside the potential after the interaction. This behaviour is very sensitive to the values of potential parameters in the model as well as to the initial conditions of a scattered soliton. Since such systems are generally non-integrable, most of the researches are still in base of numerical studies in nonlinear field theories. The collective-coordinate approach helps us to find analytical equations for the evolution of localized solutions, if one can construct such suitable variables. We will present two sets of collective-coordinate variables which are extracted from different hypotheses \cite{i13}. These help us to talk about the validity of their results and predictions.

Therefore, two models for the NKG field in a space dependent potential are presented in section 2. The two analytical models are introduced and will be discussed in section 3. The results of the two analytical models are compared for potential-barrier and potential-well systems in section 4. In section 5, we will compare our analytical results with direct numerical solutions of the equations. Some conclusions and remarks will be presented in section 6.
   
\section{Two Analytical Models for ‘the NKG Soliton-Potential’ System} \setcounter{equation}{0}

\quad\, \textbf{Model 1}. The Lagrangian of the NKG model in (1+1) dimensions is defined as 
\begin{eqnarray} \label {lag}
{\cal L}=\frac{1}{2}\partial_ {\mu}\phi\partial^{\mu}\phi-U\left (\phi\right ), 
\end{eqnarray}
where $U\left(\phi\right)$  is the potential of the field which is defined by
\begin{equation}\label{uphi}
U\left (\phi\right )=\lambda(x)\left(\frac{1}{2} \phi^2-\frac{1}{2} \phi^{4} +\frac{1}{8} \phi^{6}\right).   
\end{equation}
where $\lambda(x) =1+V(x)$. $V(x)$ is a potential parameter and carries the effects of the external potential. Potential $V(x)$ is a localized function which is nonzero only in a certain region of space. The equation of motion for the Lagrangian (\ref{lag}) is
\begin{equation}\label{eq}
\partial_ {\mu}\partial^{\mu}\phi+\lambda(x)\left(\phi-2\phi^3+\frac{3}{4} \phi^5\right)=0.  
\end{equation}
This equation cannot be solved analytically because the potential has a spatial dependence. If we take $V(x)=0$, we have the  nonlinear Klein Gordon equation with the following one-soliton solution \cite{i12}:
\begin{equation}\label{kg 1s}
\phi(x,t)={\left[1+\tanh\left(\frac{x-X(t)}{\sqrt{1-\dot{X}^{2}}}\right)\right]}^{1/2},   
\end{equation}
where $X(t)=x_0-\dot{X}t$. Here, $x_0$ and $\dot{X}$ are the soliton\textquoteright s initial position and (constant) velocity, respectively.
It is a kink-like solution as figure 1 shows. 
\begin{figure}[htbp]
\begin{center}
\leavevmode \epsfxsize=9cm \epsfbox {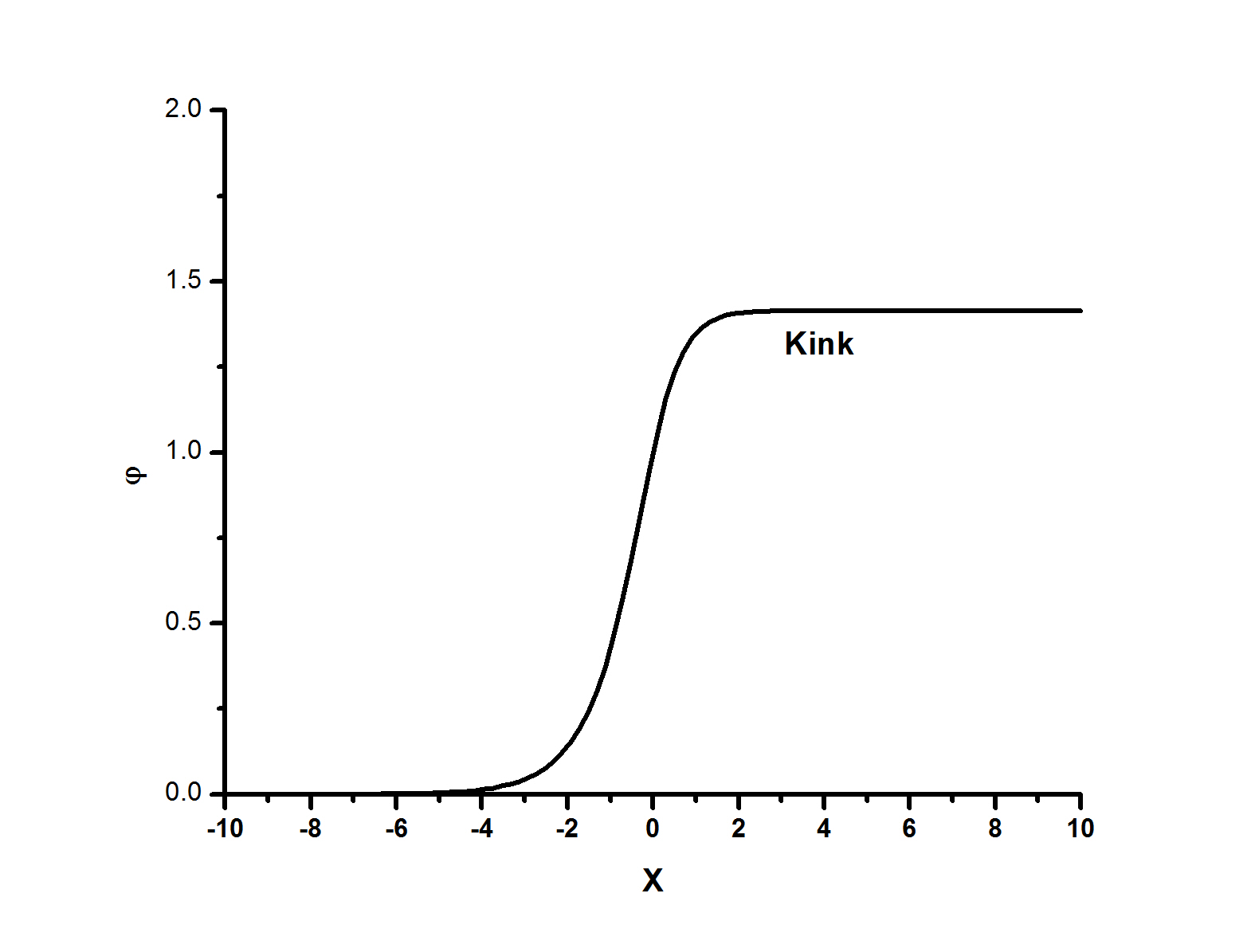}
\end {center}
\caption{Kink-like solution of the NKG described by equation (\ref{eq}) for $ x_{0}= 0$ and $\dot{X}= 0.5$ at $t=0$.}
\label{fig1}
\end{figure}
We intend to investigate the behaviour of kink solution (\ref{kg 1s}) during the interaction with an external potential using collective-coordinate technique. \

The derivation of the collective action for the motion of the vortex centers starts with the elegant idea of Manton \cite{i16}. A collective action can be constructed by substituting the collective-vortex ansatz for the field configuration with vortices at $X_{i}(t), i = 1, . . . ,N,$ into the effective field-theory action and reduce the action to a function of the collective coordinates, $L[X_{i}(t)]=\int{\cal L}\left(\psi\left(x,t,X_{i}(t)\right)\right)$ \cite{i17}. It is clear that (\ref{kg 1s}) is not the exact solution for the equation (\ref{eq}). But it is approximate solution if $V(x)$ is a local weak perturbation \cite{i15}.

By inserting the solution (\ref{kg 1s}) in the Lagrangian (\ref{lag}) and using adiabatic approximation \cite{i2} we have
\begin{align}
{\cal L}=\frac{\left(\dot{X}^{2}-1\right)\mathrm{sech}^{4}\left(x-X\right)}{8\left(1+\tanh\left(x-X\right)\right)} 
-\frac{\lambda(x)}{8}\mathrm{sech}^{2}\left(x-X\right)\left(1-\tanh\left(x-X\right)\right).
\label{lag 1}
\end{align}
In the adiabatic approximation one would suppose that the soliton velocity changes in an adiabatic process. Therefore, the soliton speed changes very slowly. Moreover, we have considered solitons with a small and slowly varying velocity. 

\textbf{Model 2}. The general form of the action in an arbitrary metric is
\begin{equation}\label{S}
S=\int{{\cal L}(\phi , \partial_{\mu}\phi)\sqrt{-g}d^{n}x dt },
\end{equation}
where "g" is the determinant of the metric $g_{\mu \nu} (x)$. Therefore, we have an effective Lagrangian ${\cal L}_{eff}={\cal L} \sqrt{-g}$. The energy density of the system can be calculated by varying both the field and the metric \cite{i6}. For the Lagrangian of the form (\ref{lag}) the equation of motion becomes \cite{i7,i13}
\begin{equation}\label{Em}
\frac {1}{\sqrt{-g}}\left (\sqrt{-g}\partial_{\mu}\partial^{\mu}\phi+\partial_{\mu}\phi\partial^{\mu}\sqrt{-g}\right )+\frac {\partial U(\phi)}{\partial \phi}=0.
\end{equation}
A space-dependent potential can be added to the Lagrangian of the system by introducing a suitable nontrivial metric for the background space-time \cite{i6,i13}. In other words, the metric carries the information about the potential. In the presence of a weak potential $V(x)$ the suitable metric is \cite{i6,i7,i13}
\begin{equation}\label{metric}
g_{\mu \nu}(x)\cong\left(
\begin{array}{clrr} 1+V(x) & 0 \\ 0 & -1
\end{array}\right).
\end{equation}
By inserting the solution (\ref{kg 1s}) in the effective Lagrangian (${\cal L}_{eff}$) with the potential (\ref{uphi}) of the NKG and using the metric (\ref{metric}), with adiabatic approximation \cite{i2,i3}, we have
\begin{eqnarray}\label{le1}
{\cal L}_{eff}&=&\sqrt{1+V(x)}\left(\left(1-V(x)\right)\dot{X}^{2}-1\right)\frac{\mathrm{sech}^{4}\left(\left(x-X\right)\right)}{8\left(1+\tanh\left(x-X\right)\right)}\quad\notag\\
&-&\frac{\sqrt{1+V(x)}}{8}\mathrm{sech}^{2}\left(x-X\right)\left(1-\tanh\left(x-X\right)\right).
\end{eqnarray} 
For the weak potential $V(x)$ (\ref{le1}) becomes
\begin{eqnarray}\label{le2}
{\cal L}_{eff}&\cong &\left(\left(1-\frac{V(x)}{2}\right)\dot{X}^{2}-\left({1+\frac{V(x)}{2}}\right)\right)\frac{\mathrm{sech}^{4}\left(\left(x-X\right)\right)}{8\left(1+\tanh\left(x-X\right)\right)}\notag\\
&-&\frac{{1+\frac{V(x)}{2}}}{8}\mathrm{sech}^{2}\left(x-X\right)\left(1-\tanh\left(x-X\right)\right).
\end{eqnarray}


\section{Collective Coordinate for the two Models}
\quad\, The Lagrangian density of the soliton is described by (\ref{lag 1}) in model 1 and (\ref{le2}) in model 2. These two equations are different in kinetic and also potential terms. We will compare them later. The soliton internal structure can be ommited by integrating the Lagrangian density (or Hamiltonian density) with respect to the variable x. The integrated Lagrangian is called collective Lagrangian. After the integration, the soliton appears as a point-like particle; however, the effect of its extended nature still reflects in the kinetic and also potential parts of the collective Lagrangian. The dynamics of the point-like particle can be described by equations which are derived from collective Lagrangian. It is interesting to compare the results of the collective equations with those of direct numerical simulation of the main Lagrangian density. Let us derive the collective Lagrangian and the point-like particle equation of motion in the two models.
     
\textbf{Model 1}. By integrating Lagrangian (\ref{lag 1}) over the variable $ x $ , $ X(t) $  remains as a collective coordinate.  If we take the potential $ V(x)=\epsilon\delta(x) $,  collective Lagrangian is derived from (\ref{lag 1}) as 
\begin {equation} \label {lap}
L=\frac{1}{4}\dot{X}^{2}-\frac{\epsilon}{8}\mathrm{sech}^{2}\left(X\right)\left(1+\tanh\left(X\right)\right)-\frac{1}{2},
\end {equation}
where $ M_0=\frac{1}{2} $  is the rest mass of the soliton in this model. The effective potential comes from equation (\ref{lap}) which is given by
\begin{equation}\label{uphi1}
U\left (\phi\right )=\frac{\epsilon}{8}\mathrm{sech}^{2}\left(X\right)\left(1+\tanh\left(X\right)\right)+\frac{1}{2}.   
\end{equation}
The NKG solitons have a special and interesting situation which has not been observed in other fields. If we plot the effective potential as a function of collective position (X), we will find that it has a spatial shift with respect to the origin. Figure 2(a) shows the effective potential as a function of position (X) for $ \epsilon=2 $. Simulations also confirm this finding for the NKG model. Figure 2(b) presents the shape of the potential barrier as seen by the soliton which has been plotted using numerical calculation.  It is a special characteristic for the NKG field theory. The source of this distinction with respect to other fields needs further consideration.
\begin{figure}[htbp]
  \begin{center}
  \leavevmode
 \epsfxsize=7cm   \epsfbox{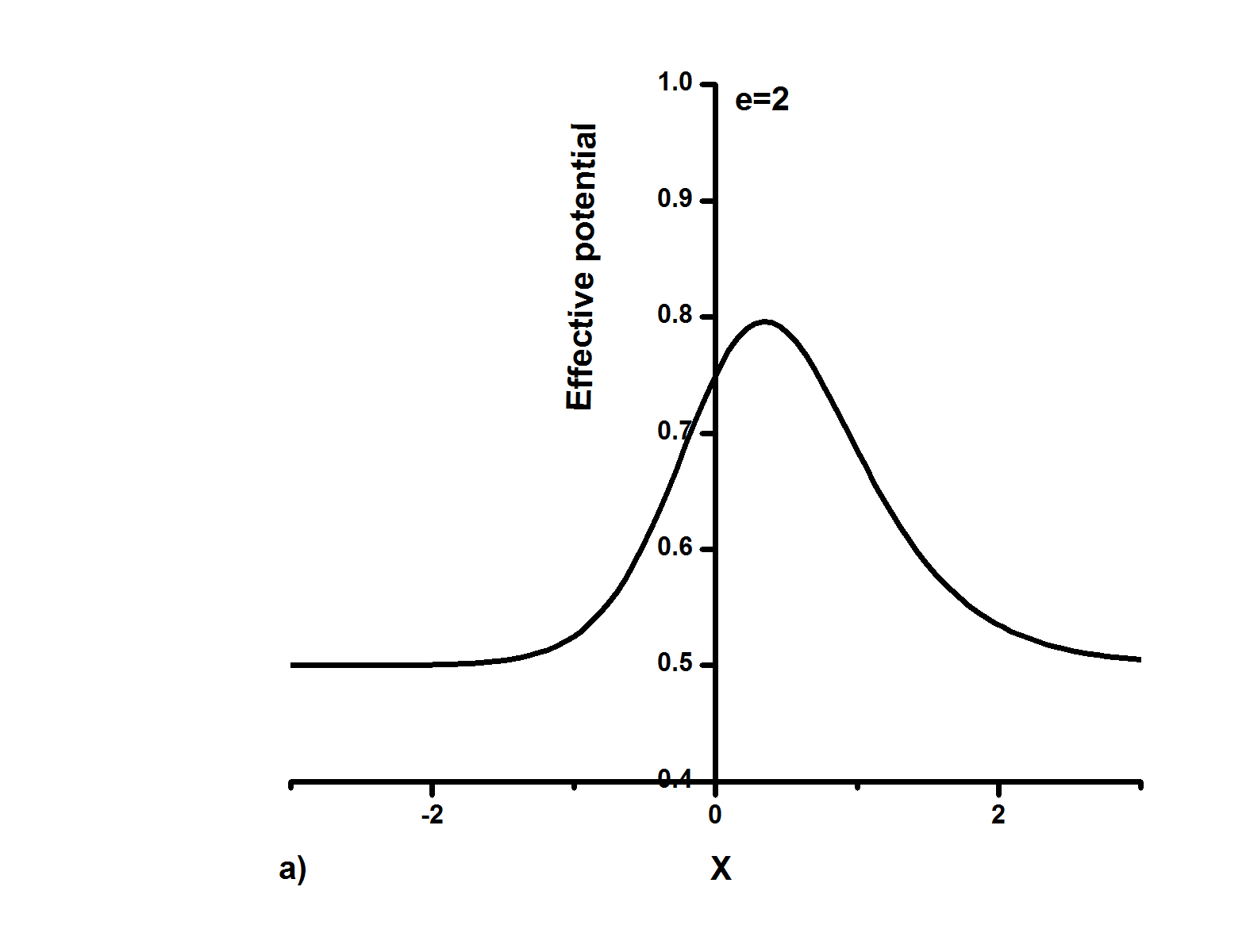}\epsfxsize=7cm \epsfbox{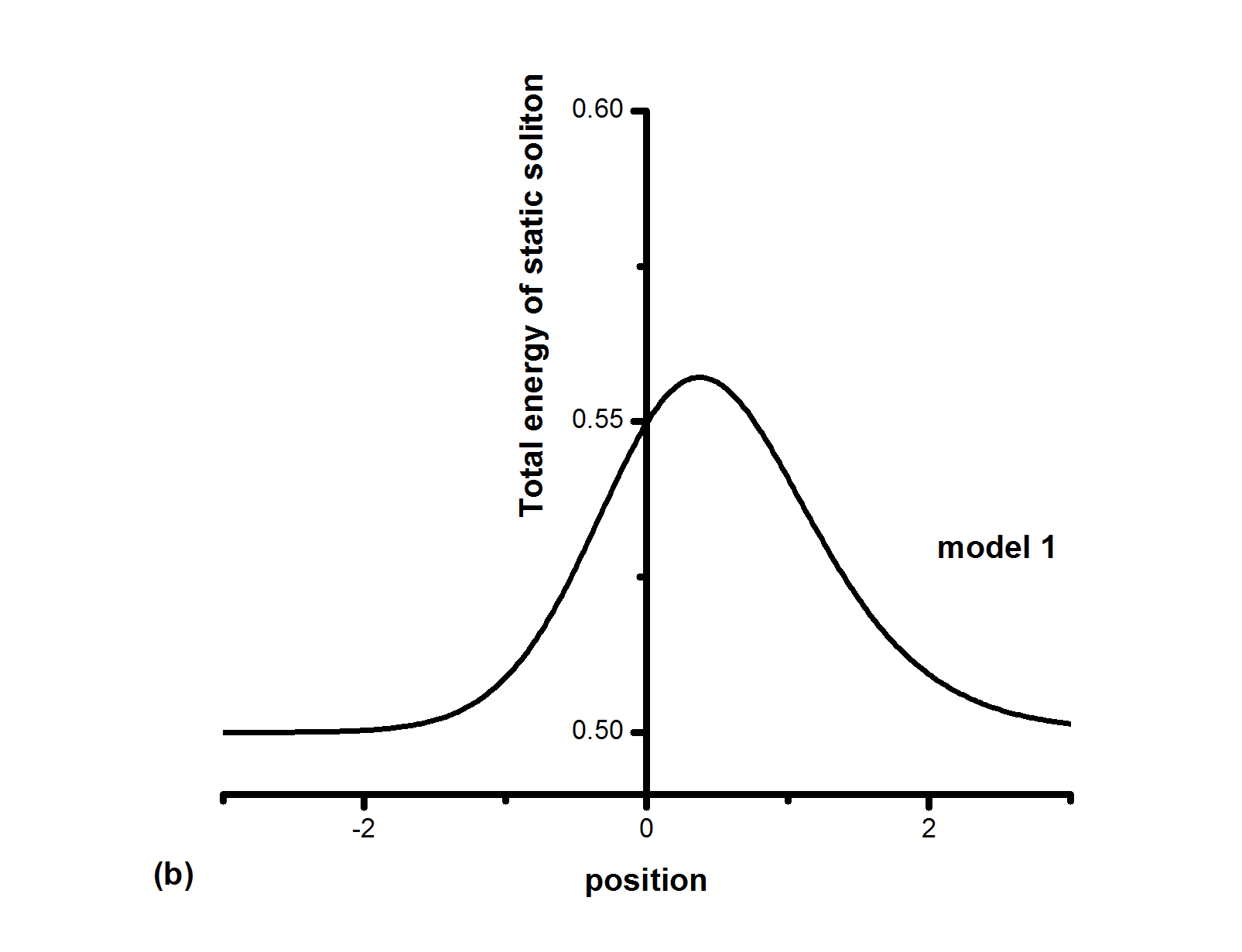} 
  \end{center}
  \caption{(a) Effective potential as a function of position with $\epsilon=2$. (b) Potential barrier as seen by the soliton in model 1 using numerical simulations.}
  \label{fig:fig2}
\end{figure}

The equation of motion for the variable $ X(t) $ is derived from (\ref{lap}) as
\begin{equation} \label{lap1}
\frac{1}{2}\ddot{X}-\frac{\epsilon}{2}\mathrm{sech}^{2}\left(X\right)\left[\frac{3}{4}{\tanh^{2}\left(X\right)}+\frac{1}{2}\tanh\left(X\right)-\frac{1}{4}\right]=0.
\end{equation}
We can define a collective force on the soliton if we look at the above equation as $ F=M\ddot{X} $, where M is the rest mass of the soliton. Therefore, we have   
\begin{equation} \label{force}
F=\frac{\epsilon}{2}\mathrm{sech}^{2}\left(X\right)\left[\frac{3}{4}{\tanh^{2}\left(X\right)}+\frac{1}{2}\tanh\left(X\right)-\frac{1}{4}\right].
\end{equation}
The above equation shows that the peak of the soliton moves under the influence of a complicated force which is a function of an external potential and soliton position.
Suppose that a soliton moves toward a potential barrier. Its velocity will reduce due to the effect of a repulsive force. While the soliton is moving away, its velocity increases. Figures 3(a) and 3(b) show the force exerted by the potential well and barrier on the soliton for $ \epsilon=-4 $ and $ \epsilon=4 $ respectively. This is in agreement with the observed behaviour for the NKG model as shown in figure 2. These figures also show that the center of the force is not located in the origin.
\begin{figure}[htbp]
  \begin{center}
    \leavevmode
 \epsfxsize=7cm   \epsfbox{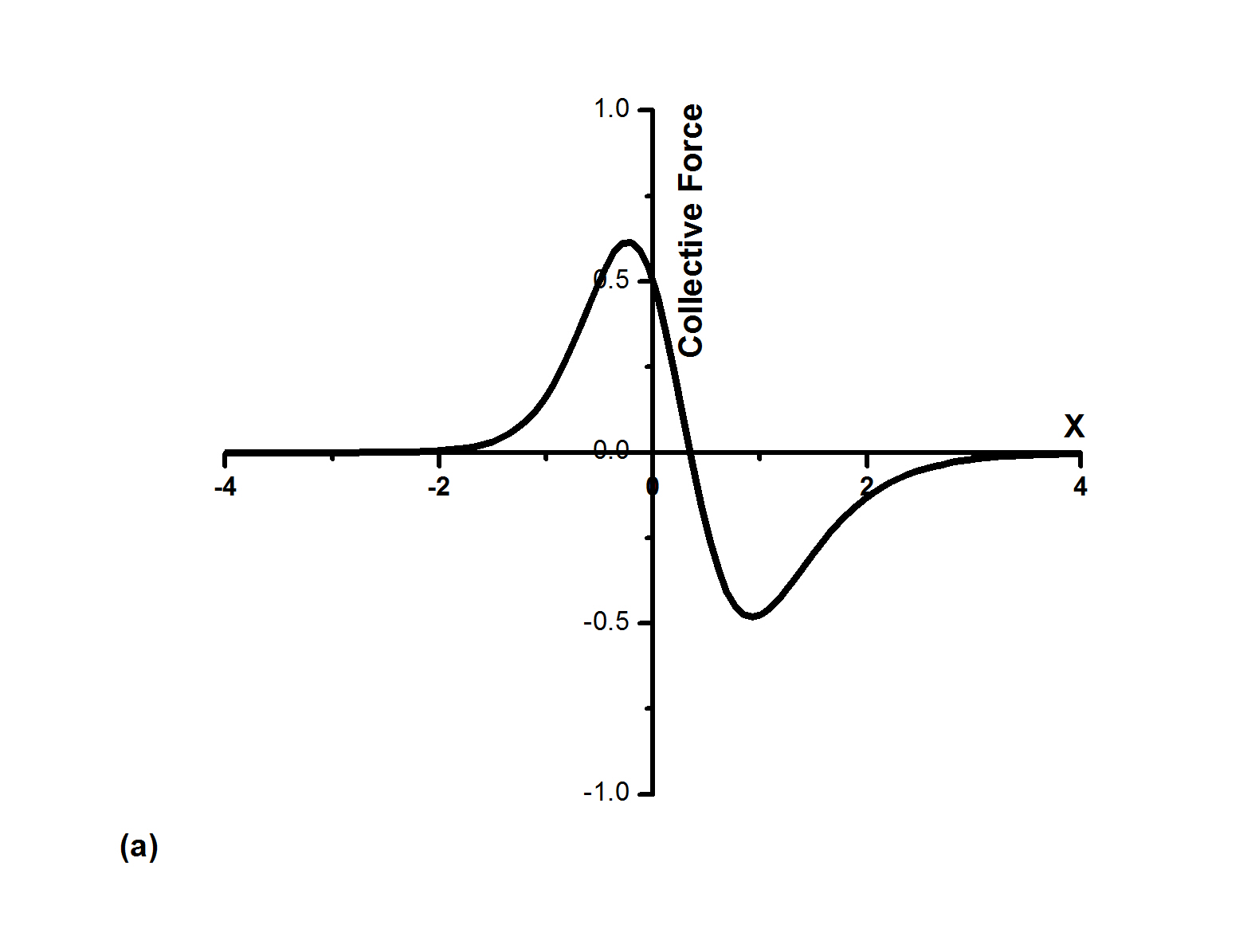}\epsfxsize=7cm \epsfbox{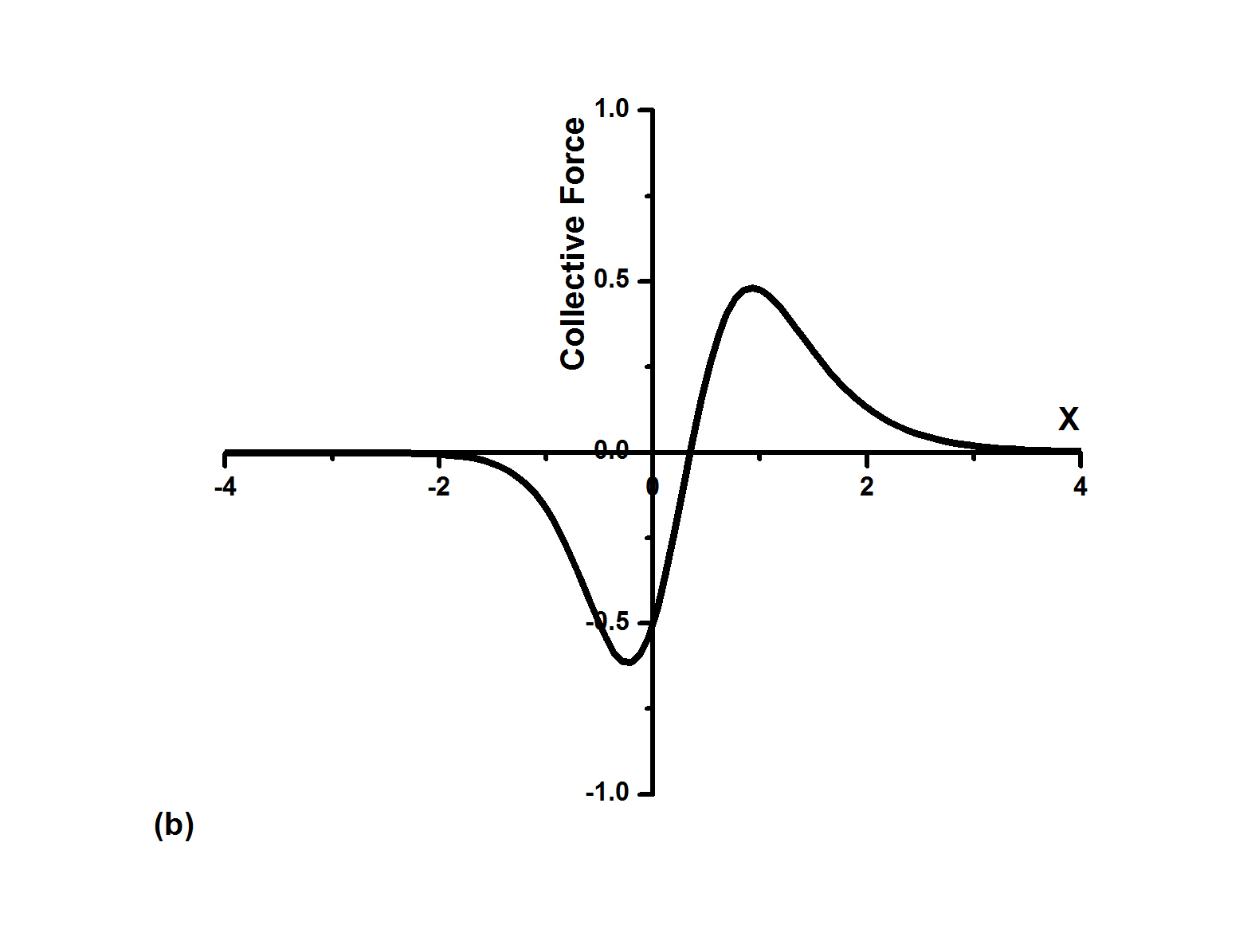} 
  \end{center}
  \caption{(a) The force on the soliton by a potential well with $\epsilon=-4$. (b) The force on the soliton by a barrier with $\epsilon=4$.  }
  \label{fig:fig3}
\end{figure}

Interestingly, equation (\ref{lap1}) has an exact solution for $\dot{X}$ as follows
\begin{eqnarray} \label{lap2}
\dot{X}^{2}-\dot{X_0}^{2}&=&\frac{\epsilon}{2}[{\tanh^{3}\left(X\right)}+\tanh^{2}\left(X\right)-\tanh\left(X\right)\notag\\
&-&\tanh^{3}\left(X_0\right)-\tanh^{2}\left(X_0\right)+\tanh\left(X_0\right)],
\end{eqnarray}
where $ X_0 $  and $ \dot{X_0} $  are the soliton\textquoteright s initial position and initial velocity, respectively. Some of the physical features of a soliton-potential system can be discovered using equation (\ref{lap2}). Collective energy is obtainable from Lagrangian (\ref{lap}) as follows
\begin{equation} \label{energy1}
E=\frac{1}{4}\dot{X}^{2}+\frac{\epsilon}{8}\mathrm{sech}^{2}\left(X\right)\left(1+\tanh\left(X\right)\right)+\frac{1}{2}.
\end{equation}
It is the energy of a particle with the mass of $ M_0=\frac{1}{2} $  and the velocity of $ \dot{X} $,  where the particle moves under the influence of an external effective potential. By substituting $\dot{X}$ from (\ref{lap2}) in the equation (\ref{energy1}), it is shown that the energy of the system is a function of soliton\textquoteright s initial conditions $ X_0 $  and $ \dot{X_0} $ and therefore it is conserved. 

\textbf{Model 2}. The above results can be calculated using model 2. For the Lagrangian (\ref{le2}) $ X(t) $ remains  a collective coordinate if we integrate (\ref{le2}) over the variable $x$. If we take the potential $V(x)=\epsilon\delta(x) $, the collective Lagrangian becomes
\begin{align} \label{L1}
L=\left(\frac{1}{4}-\frac{\epsilon \mathrm{sech}^{4}\left(X\right)}{16\left(1-\tanh\left(X\right)\right)}\right)\dot{X}^{2}
-\frac{\epsilon}{8}\left(\frac{\mathrm{sech}^{4}(X)}{1-\tanh\left(X\right)}\right)-\frac{1}{2}.
\end{align}
The equation of motion for the variable $X(t)$ is derived from (\ref{L1}):
\begin{align} \label{eom}
\left(\frac{1}{2}-\frac{\epsilon \mathrm{sech}^{4}\left(X\right)}{8\left(1-\tanh\left(X\right)\right)}\right)\ddot{X}-\frac{\epsilon}{16}\mathrm{sech}^{2}\left(X\right)\qquad\qquad\notag\\
\times\left(-3\tanh^{2}\left(X\right)-2\tanh\left(X\right)+1\right)\left(\dot{X}^2-2\right)=0.
\end{align}
The above equation describes the soliton trajectory which moves under the influence of a collective force. The collective force is a function of the soliton\textquoteright s position and velocity. $\dot{X}(t)$ can be calculated as a function of $X(t)$ by integrating (\ref{eom}) as follows
\begin {equation} \label {eom 1}
\frac{\dot{X}^2-2}{\dot{X_0}^2-2}=\frac{1-\frac{\epsilon \mathrm{sech}^{4}\left(X_0\right)}{4\left(1-\tanh\left(X_0\right)\right)}}{1-\frac{\epsilon \mathrm{sech}^{4}\left(X\right)}{4\left(1-\tanh\left(X\right)\right)}},
\end {equation}
where $ X_0 $  and  $ \dot{X_0} $ are the soliton\textquoteright s initial position and initial velocity, respectively. The energy of the soliton in the presence of the potential $ V(x)=\epsilon\delta(x) $ using model 2 becomes
\begin{align} \label{energy2}
E=\left(\frac{1}{4}-\frac{\epsilon \mathrm{sech}^{4}\left(X\right)}{16\left(1-\tanh\left(X\right)\right)}\right)\dot{X}^{2}
+\frac{\epsilon}{8}\left(\frac{\mathrm{sech}^{4}(X)}{1-\tanh\left(X\right)}\right)+\frac{1}{2}.
\end{align}
Equation (\ref{energy2}) shows that the rest mass is a function of the soliton\textquoteright s position in model 2. This is the source of some differences between the two models which will be discussed in the next section. In section 4, some features of the soliton-potential dynamics are studied analytically using equations (\ref{eom 1}) and (\ref{energy2}).
\section {Comparing the Models}
\quad\, \textbf{Potential barrier}. There are two different trajectories for a soliton during the interaction with an effective potential barrier which depend on the soliton\textquoteright s initial conditions. A soliton with a low velocity reflects back from the barrier and a high-velocity soliton climbs up the barrier and passes over it. So, these two situations can be distinguished by a critical velocity. The total energy of the soliton-potential is conserved in both models as mentioned before. Therefore, we can find the critical velocity with a simple analysis of the energy of the soliton without any numerical simulations. Both equations (\ref{energy1}) and (\ref{energy2}) are reduced to $ E(X=\infty)=\frac{1}{4}\dot{X_0}^{2}+\frac{1}{2} $ when the soliton is far from the center of the delta-like potential which is located at the origin. It is the energy of a particle with the mass of  $M_{0}= \frac{1}{2} $ and velocity of $\dot{X_0}$.  The energy of a soliton in the origin ($ X=0 $)  comes from (\ref{energy1}) and (\ref{energy2}) for the two models:  $ E_{1}(X=0)=\frac{1}{4}\dot{X_0}^{2}+\frac{\epsilon}{8}+\frac{1}{2} $  for model 1 and  $ E_{2}(X=0)=\left(\frac{1}{4}-\frac{\epsilon}{16}\right)\dot{X_0}^{2}+\frac{\epsilon}{8}+\frac{1}{2} $  for model 2. The minimum energy of soliton in this position is  $ E=\frac{\epsilon}{8}+\frac{1}{2} $  for the two models. On the other hand, a soliton which comes from the infinity with initial velocity $ v_c $  has the energy of $ E=\frac{1}{4}{v_c}^{2}+\frac{1}{2} $. So it is easy to calculate the critical velocity of soliton by comparing the energy of the soliton at the origin to its energy at infinity. The critical velocity is calculated as $ v_c=\sqrt{\frac{\epsilon}{2}} $ using both models. The same result is derived by substituting $ \dot{X}=0 $ , $ \dot{X_0}=v_c $ , $ {X_0}=\infty $  and $ X=0 $  in (\ref{lap2}) and (\ref{eom 1}).

Note that the critical velocity of a soliton depends on its initial position as well as its initial velocity.	For a soliton which is located at some position like $ X_0 $  (which is not necessarily infinity) the critical velocity will not be $ v_c=\sqrt{\frac{\epsilon}{2}} $. So a soliton in the initial position $ X_0 $  with initial velocity of $ \dot{X_0} $  has the critical initial velocity if its velocity becomes zero at the top of the barrier $ X=0 $. Consider a soliton with initial conditions of  $ X_0 $ and $ \dot{X_0} $. If we set $ X=0 $ and  $ \dot{X}=0 $ in equations (\ref{lap2}) and (\ref{eom 1}) then $ v_c=\dot{X_0} $. Thus for model 1 we have
\begin {equation} \label {cv1}	
v_{c}=\sqrt{-\frac{\epsilon}{2}\left(\tanh\left(X_0\right)-\tanh^{2}\left(X_0\right)-\tanh^{3}\left(X_0\right)\right)}.
\end {equation}
But the critical velocity in model 2 becomes
\begin {equation} \label {cv2}	
v_{c}=\sqrt{\frac{2\epsilon\left(1-\tanh(X_0)-\mathrm{sech}^{4}(X_0)\right)}{4-4\tanh(X_0)-\epsilon \mathrm{sech}^{4}(X_0)}}.
\end {equation} 
Figure 4 shows the critical velocity as a function of the potential strength for $ X_0=-1 $ in the two models. This figure shows that model 1 predicts smaller critical velocity compared to model 2 due to the differences between rest masses of these models. It is clear that a soliton with a great rest mass needs smaller velocity to reach the potential peak. The difference between the rest mass of the soliton in model 1 and that of the soliton in model 2 can be calculated using (\ref{energy1}) and (\ref{energy2}) as
\begin {equation} \label {Ek}	
\triangle E_{kinetic}=\frac{\epsilon \mathrm{sech}^{4}(X)}{16\left(1-\tanh(X)\right)}\dot{X}^2.
\end {equation} 
The above equation shows that the two models predict equal rest mass at the infinity. But the difference between the calculated rest mass in the two models increases as the strength of the potential increases.  Figure 4 shows this phenomenon explicitly.
\begin{figure}[htbp]
\begin{center}
\leavevmode \epsfxsize=9cm \epsfbox {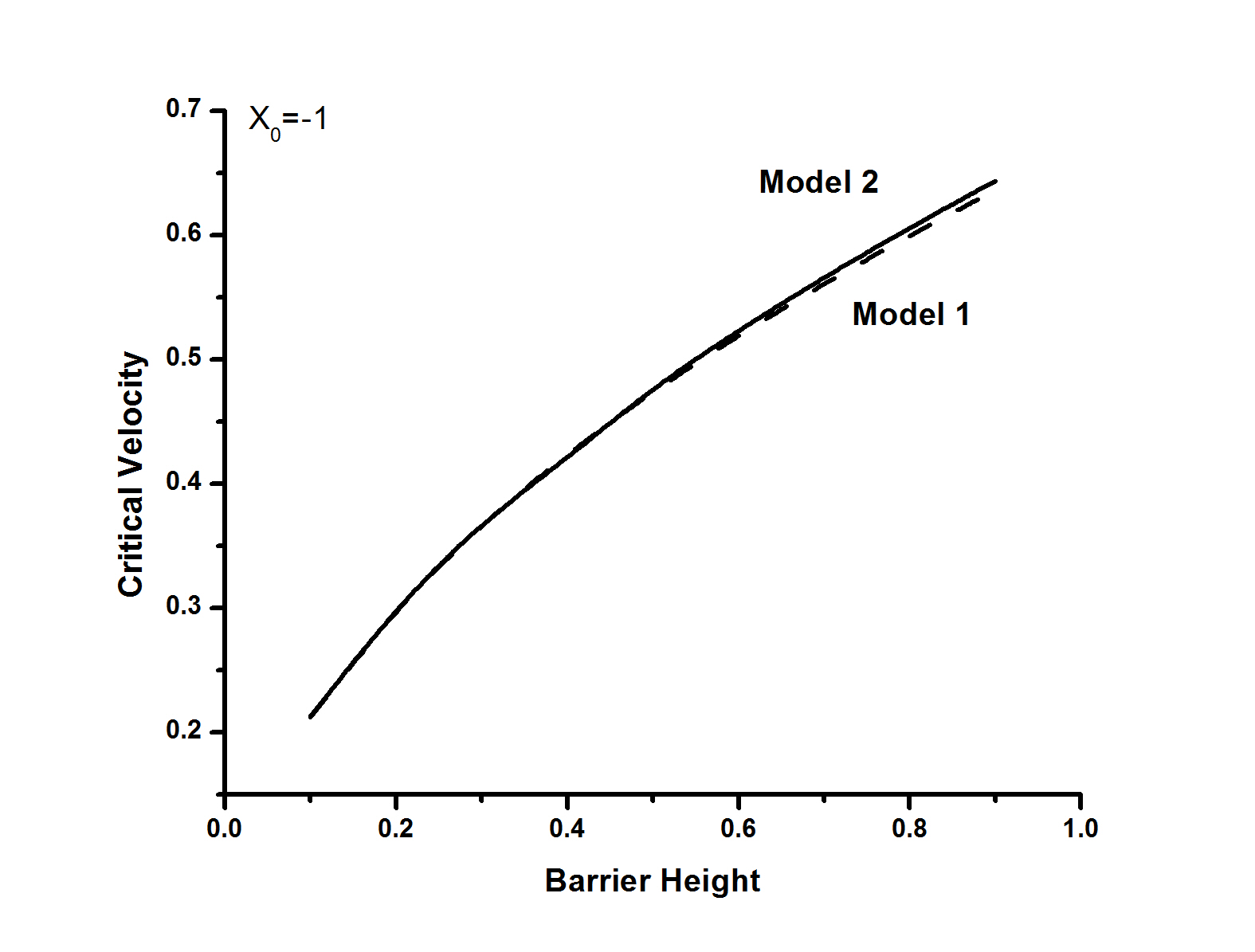}
\end {center}
\caption{Critical velocity as a function of barrier height in both models for initial position $ X_0=-1 $.}
\label{fig4}
\end{figure}
 It is interesting to depict the critical velocity as a function of initial position. The critical velocity has been demonstrated as a function of the initial position in figure 5 for the two models with $ \epsilon=0.5 $. This figure shows a considerable agreement between the two models. For a soliton at infinity, the two models demonstrate confirming results as shown in figure 5. This figure also demontrates that a soliton needs lower initial velocity to pass over the barrier if it is closer to the center of the potential.
\begin{figure}[htbp]
\begin{center}
\leavevmode \epsfxsize=9cm \epsfbox {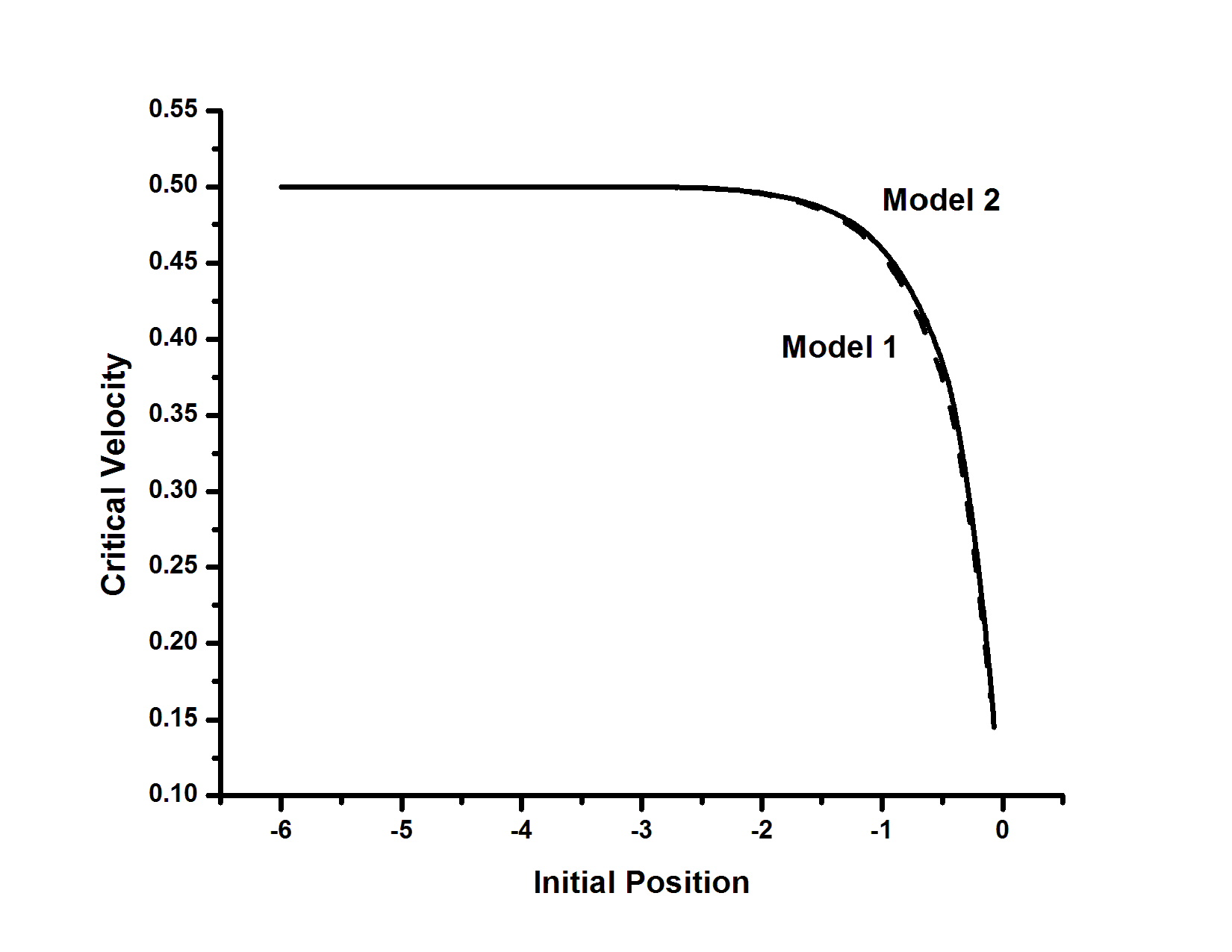}
\end {center}
\caption{Critical velocity as a function of initial position in both models for $ \epsilon=0.5 $.}
\label{fig5}
\end{figure}

\textbf{Soliton-well system}. Let us consider a soliton which moves toward a frictionless potential well. This situation is worth investigating because of some differences between a point particle and a soliton in the potential well. A point particle falls in the well with an increasing velocity and reaches the bottom of the well with its maximum speed. After that, it will climb the well with a decreasing velocity and finally passes through the well. Its final velocity, after the interaction, equals its initial velocity.

We will have a potential well by replacing $ \epsilon $ with $ -\epsilon $  in the equation (\ref{lap2}). The solution for the system in model 1 is
\begin{eqnarray}\label{well1}
\dot{X}^{2}-\dot{X_0}^{2}&=&-\frac{\epsilon}{2}[{\tanh^{3}\left(X\right)}+\tanh^{2}\left(X\right)-\tanh\left(X\right)\notag\\
&-&\tanh^{3}\left(X_0\right)-\tanh^{2}\left(X_0\right)+\tanh\left(X_0\right)].
\end{eqnarray}
Similary, for model 2 we have
\begin {equation} \label {well2}
\frac{\dot{X}^2-2}{\dot{X_0}^2-2}=\frac{1+\frac{\epsilon \mathrm{sech}^{4}\left(X_0\right)}{4\left(1-\tanh\left(X_0\right)\right)}}{1+\frac{\epsilon \mathrm{sech}^{4}\left(X\right)}{4\left(1-\tanh\left(X\right)\right)}}.
\end {equation}
We can define an escape velocity instead of a critical velocity for a soliton-well system. The escape velocity is the minimum velocity for a soliton which can pass through a well. A soliton in an initial position $ X_0 $  reaches the infinity with a zero final velocity if its initial velocity is
\begin{equation} \label{escape 1}	
\dot{X}_{escape 1}=\sqrt{\frac{\epsilon}{2}\left(\mathrm{sech}^{2}\left(X_0\right)-\tanh ^{3}\left(X_0\right)+\tanh\left(X_0\right)\right)},
\end{equation}
and
\begin{equation} \label{escape 2}	
\dot{X}_{escape 2}=\sqrt{\frac{2\epsilon \mathrm{sech}^{4}\left(X_0\right)}{4-4\tanh(X_0)+\epsilon \mathrm{sech}^{4}(X_0)}},
\end{equation}
which are calculated using models 1 and 2 respectively. In other words, a soliton which is located in the initial position $ X_0 $  can escape to infinity if its initial velocity $ \dot{X_0} $  is greater than the escape velocity $ \dot{X}_ {escape} $. Figures 6 and 7 show the escape velocities from a potential well as a function of the well depth (figure 6) and soliton\textquoteright s initial position (figure 7) using the two models. Due to its bigger rest mass, the soliton in model 2 needs lower escape velocity in comparison with the soliton in model 1.  This is obvious in the equation (\ref{Ek}).
\begin{figure}[htbp]
\begin{center}
\leavevmode \epsfxsize=9cm \epsfbox {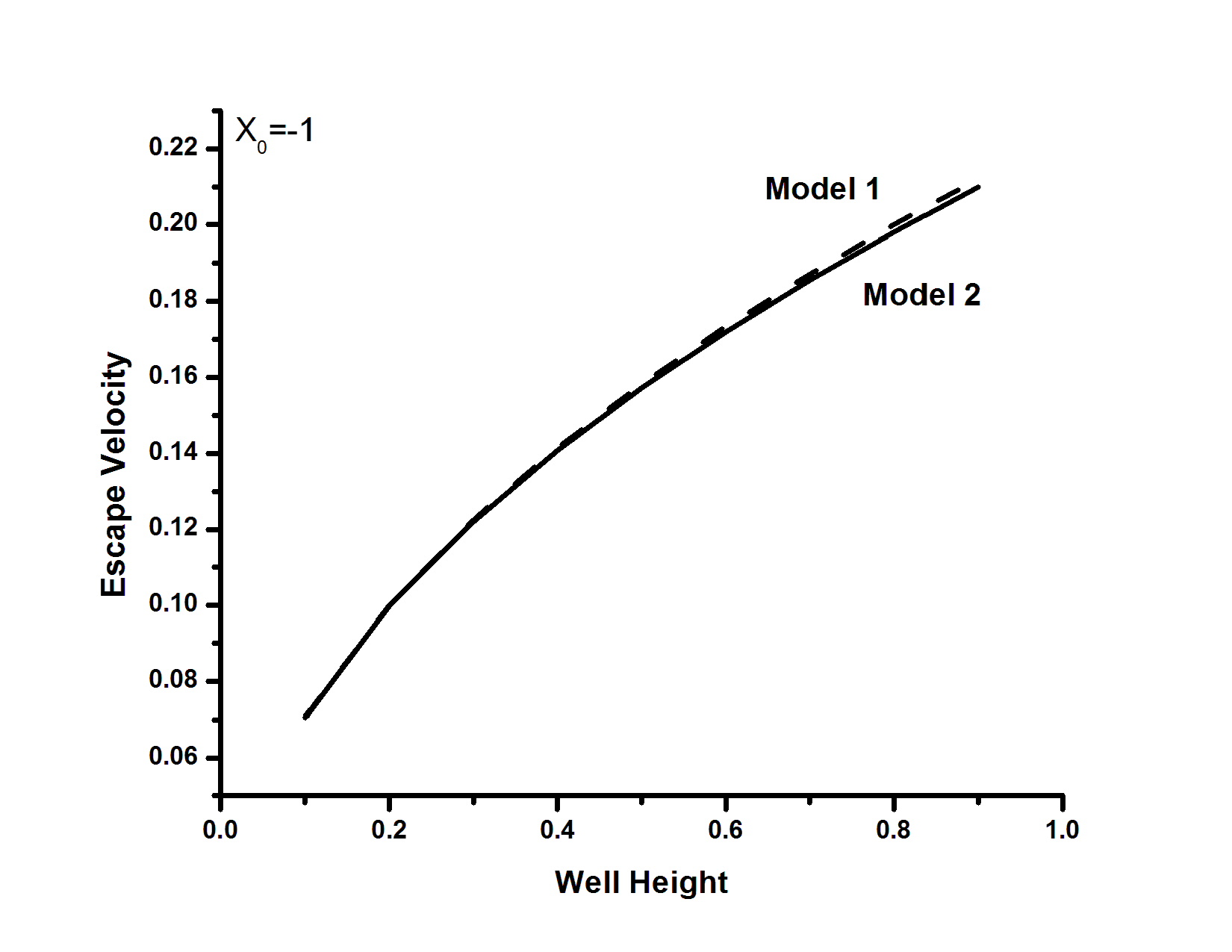}
\end {center}
\caption{Escape velocity as a function of the well depth in both models for initial position $ X_0=-1 $.}
\label{fig6}
\end{figure}

Consider a soliton which moves toward the potential well with an initial velocity $ \dot{X_0} $  smaller than the escape velocity $ \dot{X}_{escape} $. The soliton reaches a maximum distance $ X_{max} $  from the center of the potential with a zero velocity and then comes back toward the center of the potential well. Therefore, the soliton oscillates in the well with the amplitude of $ X_{max} $. The required initial velocity to reach $ X_{max} $  is found from (\ref{well2}) for model 2 as
\begin {equation} \label {initial p}	
\dot{X_0}=\sqrt{\frac{2\epsilon\left(\mathrm{sech}^{4}X_0\left(1-\tanh X_{max}\right)- \mathrm{sech}^{4}X_{max}\left(1-\tanh X_0 \right)\right)}{\left(1-\tanh X_{max}\right)\left(4-4\tanh X_0+\epsilon \mathrm{sech}^{4}X_0\right)}}.
\end {equation}
It is clear that the soliton oscillates around the well if its initial velocity is lower than the escape velocity. The period of the oscillation can be calculated numerically using equation (\ref{well2}).
\begin{figure}[htbp]
\begin{center}
\leavevmode \epsfxsize=9cm \epsfbox {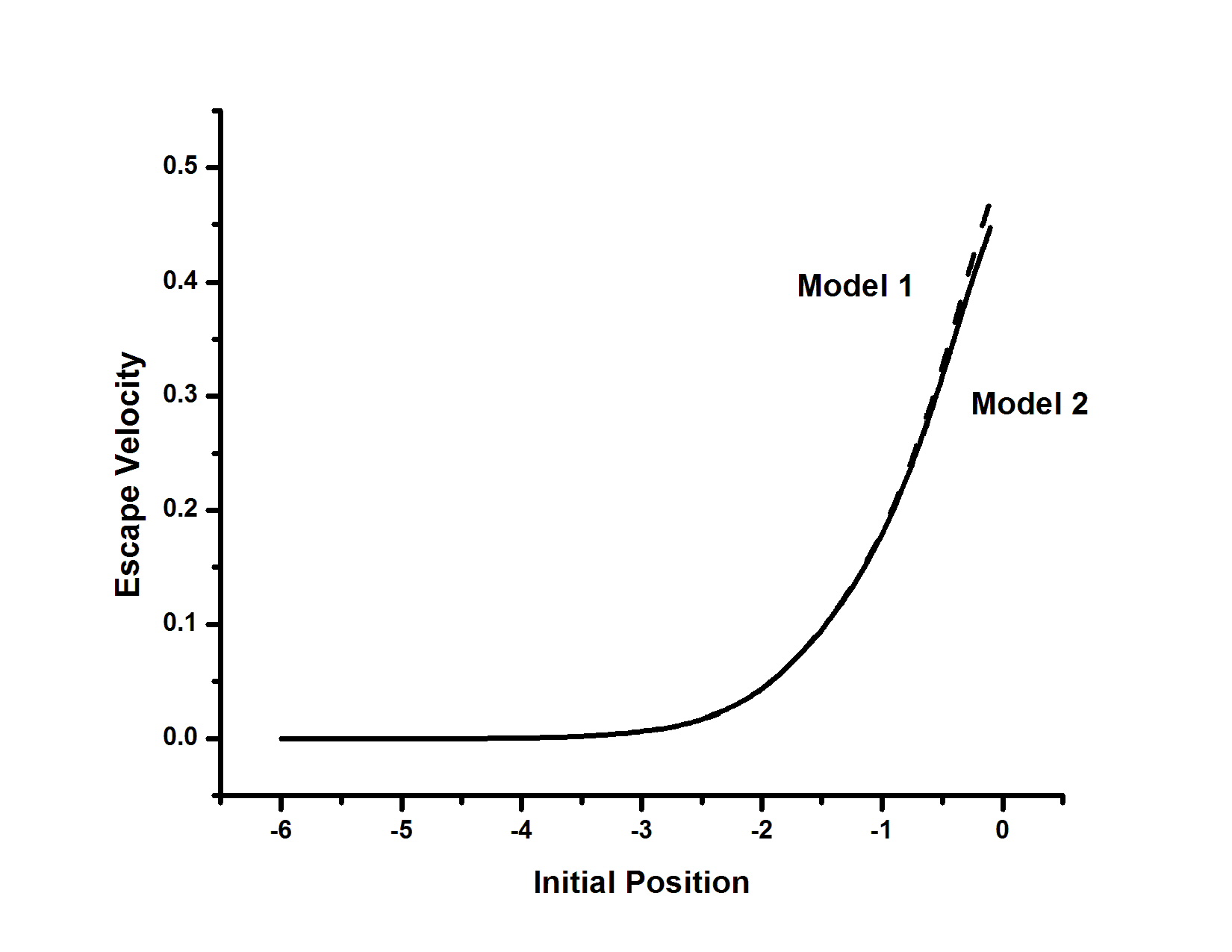}
\end {center}
\caption{Escape velocity as a function of initial position in both models for $ \epsilon=0.5 $.}
\label{fig7}
\end{figure}


\section{Analytical Results vs. Numerical Simulation}

\quad\, It is shown that the general behaviour of a soliton-potential system is almost the same in models 1 and 2. However, we can find some small differences between the dynamics of a soliton in the two models. It is important to compare the results of analytical models to direct numerical solutions. Here, we will compare the analytic results of model 1 to its numerical solution. It is clear that the same comparison can be done for model 2. The soliton equation of motion for a small potential in model 1 is \cite {i17}
\begin {equation} \label {eom2}	
\phi_{tt}-\phi_{xx}+\left(1+\sigma\delta(x)\right)\frac{\partial {U}}{\partial{\phi}}=0.
\end {equation}                                                                                                                                      
Both models (for a delta-like potential) have a parameter in their equation of motion which controls the strength of the external potential. It is possible to compare the strength parameters in a specific situation by simulating and adjusting parameters in order to have the same results in different models for that specific situation. It is expected to find approximately the same relation between the parameters in other situations. 
We found that the critical velocity for a soliton-barrier system is  $ v_c=\sqrt{\frac{\epsilon}{2}} $. It is possible to adjust the strength parameter $\epsilon$ in an analytical model with the same parameter in (\ref{eom2}) by means of $v_{c}$. Numerical simulations using equation (\ref{eom2}) show the same behaviour for a critical velocity. An effective potential can be found by interpolation of simulation results on the $ v_c=\sqrt{\frac{\epsilon_{eff}}{2}}=\sqrt{\frac{\alpha+\epsilon\beta}{2}} $. $\epsilon_{eff}$ is the potential strength (as an effective parameter) in numerical simulation while $\epsilon$ is similar parameter in model 1. $\epsilon_{eff}$ can be found by fitting the numerical results on a theoretical diagram as  
\begin {equation} \label {eff}	
\epsilon_{effective}=\left(0.0434\pm 0.01061\right)+\left(0.76462\pm 0.02479\right)\epsilon.
\end {equation}                                                                                
Figure 8 shows the result of simulations of equation (\ref{eom2}) for the NKG model. Our simulations show a very good agreement between numerical results and theoretical predictions for other features of soliton-potential interaction.  

\begin{figure}[htbp]
\begin{center}
\leavevmode \epsfxsize=9cm \epsfbox {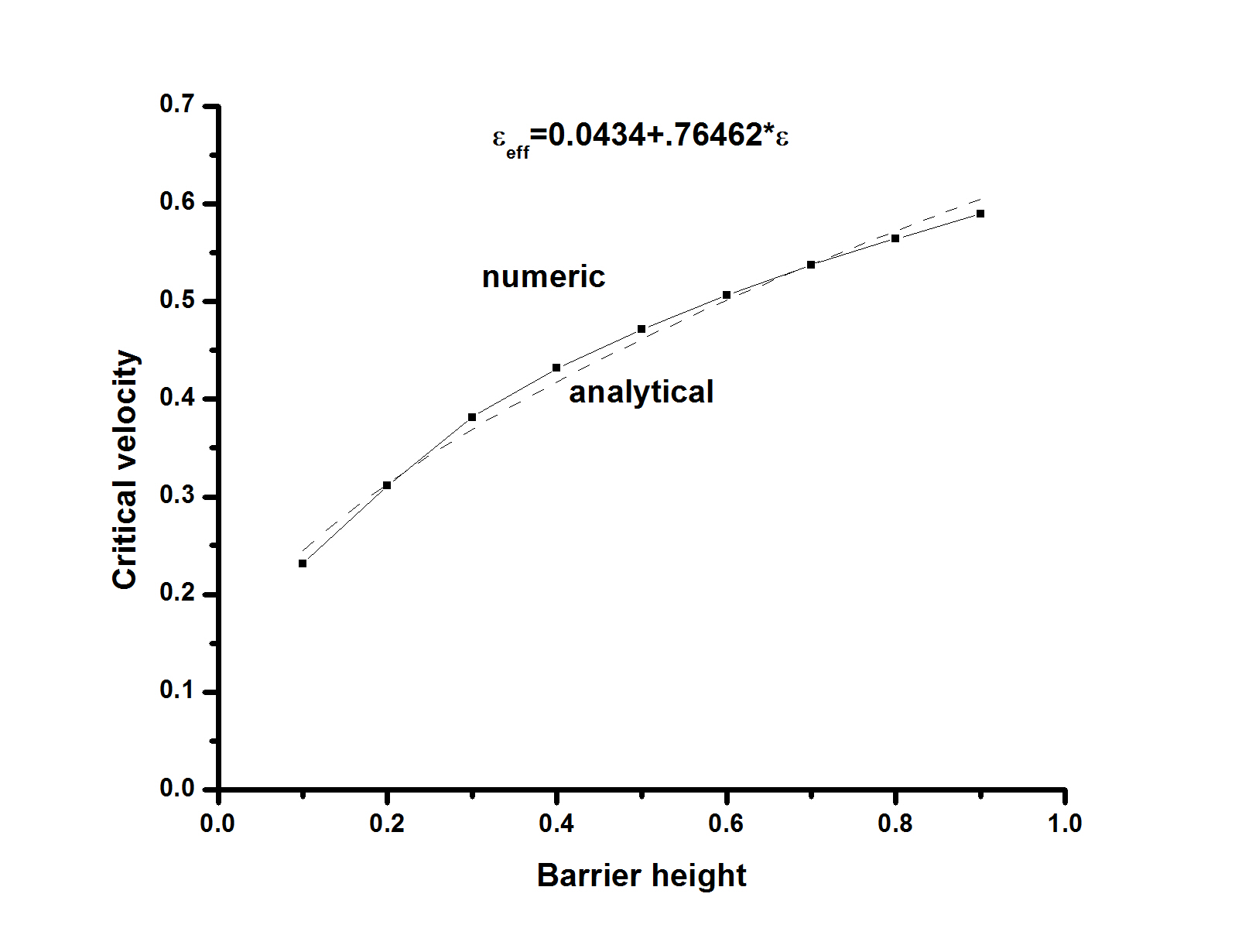}
\end {center}
\caption{Critical velocity as a function of $\epsilon$ with results of simulation using equation (\ref{eom2}) and analytical model.}
\label{fig8}
\end{figure}

\section{Conclusions and Remarks}
\quad\, Two analytical models for the interaction of the NKG solitons with delta function potential have been presented. The two models predict a critical velocity for the soliton-barrier interaction which is a function of initial conditions and the potential identities. For a soliton-well system an escape velocity has been introduced instead of the critical velocity. These models are able to explain most of the features of the system analytically. We have observed that the center of the potential, as seen by the NKG solitons, is quite different from the real position of the potential center. Numerical simulations are in agreement with theoretical predictions of the models. Our models fail to predict the narrow windows of soliton reflection from the potential well. So, it is expected to find a better model with a suitable collective-coordinate method to explain this behaviour. These models can be used to predict the soliton behaviour in the other field theories beside the NKG model.


\end{document}